\documentstyle[12pt]{article}
\begin{document}
\begin{titlepage}
\title{GRAVITATIONAL DUST COLLAPSE WITH COSMOLOGICAL CONSTANT}
\author{Mahdy Cissoko$^1$, J\'ulio C. Fabris$^2$, J\'er\^ome Gariel$^1$,\\
G\'erard Le Denmat$^1$ and
Nilton O. Santos$^{1,3}$\\
\mbox{\small $^1$Universit\'e Paris VI, CNRS/UPRESA 7065}\\
\mbox{\small Gravitation et Cosmologie Relativistes}\\
\mbox{\small Tour 22-12, 4\`eme \'etage, Bo\^{\i}te 142, 4 place
Jussieu}\\
\mbox{\small 75005 Paris, France.}\\
\mbox{\small $^2$Departamento de F\'{\i}sica, Universidade Federal do
Esp\'{\i}rito Santo}\\
\mbox{\small Vit\'oria, CEP29060-900, Esp\'{\i}rito Santo, Brazil.}\\
\mbox{\small $^3$Laborat\'orio de Astrof\'{\i}sica e Radioastronomia}\\
\mbox{\small Centro Regional Sul de Pesquisas Espaciais - INPE/MCT}\\
\mbox{\small Cidade Universit\'aria, 97105-900 Santa Maria RS, Brazil.}}
\maketitle
\begin{abstract}
We study the fate of gravitational collapse in presence of a cosmological
constant.
The junctions conditions between static and non-static space-times are
deduced.
Three apparent horizon are formed, but only two have physical
significance,
one of them being the black hole horizon and the other the cosmological
horizon.
The cosmological constant term slows down the collapse of matter, limiting
also the size of the black hole.
\end{abstract}
\end{titlepage}
\section{Introduction}
The possible existence of a cosmological constant is one of the most
important challenges
in high energy physics today \cite{Weinberg}. Quantum field theory
predicts
the existence of a huge
vacuum energy density, of order of $\rho_V \sim 10^8 GeV^4$ (in units
where $\hbar = c = 1$). Introducing it in the
Einstein's field
equation, this would lead to a overclosed Universe. To avoid this problem,
we can include
a cosmological constant term in the Einstein's equation, that would cancel
exactly this
contibution coming from the quantum vacuum. Of course, we must fine tune
the
constant
term introduced by hand, presenting esthetical and conceptual drawbacks.
\par
However, a surprising recent result coming from the analysis of high
redshift
supernovae, indicates that the Universe may be accelerating now
\cite{Cohn,Riess,Perlmutter}. This suggests
that there is in fact a cosmological constant, that dominates the content
of
energy of the Universe today. The cosmological implications of the
existence
of a cosmological constant today (we must remember that the inflationary
scenario
makes use extensively of a cosmological constant term in the primordial
Universe)
are enormous, concerning not only the evolution of the Universe but also
the
structure formation and age problems.
\par
In any way, it is reasonable to think that under extreme physical
conditions
quantum effects may play an important role even in gravitating systems and
the cosmological constant problem comes, under this circumstances, to
surface.
The gravitational collapse is one example of these extreme physical
conditions
where black holes seem to be formed. If we study dynamically the formation
of a black
hole, through a spherically symmetric spacetime in presence of dust
matter,
the matter suffers a gravitational collapse process, leading first to the
formation
of an apparent horizon, and then of a singularity. The mass, and
consequently the dimension,
of the black hole has not, in principle, any restrictions.
\par
The goal of this work is to extend the study of the gravitational collapse
of dust
matter in the presence of a cosmological constant. This study has been
made
using
static configuration for the spherically symmetric
space-time\cite{Garfinkle,Hayward}. We extend this
analysis through the introduction of dynamics, considering the collapse
process itself.
Some of the results already found, like the limitation in size of a
possible
black hole,
are reobtained. However, we can analyze dynamically the appearence of two
physical
apparent horizons, the slow down of the collapse process, giving some
simple
models
to understand why the cosmological constant, due to its repulsive
character,
limits
the existence of possible black holes.
\par
This article is organized as follows. In the next section, we consider the
junctions
conditions between a static and a non-static spherically symmetric
spacetimes.
In section 3, we solve the field equations using the Tolman-Bondi metric,
with
a matter content composed of dust and a cosmological constant.
In section 4, we specialize the solution obtained in the previous section.
The apparent horizons
in such spacetimes are analyzed in section 5, and the consequences of the
presence of a cosmological constant are considered. We end with the
conclusions
where we analyze and
try to interpret our results.

\section{Junction conditions}

The object of this section is to derive the conditions to be satisfied at
the surface of a collapsing perfect fluid sphere with cosmological
constant.
We assume spherical symmetry about an origin 0. Inside a spherical surface
$\Sigma$, center 0, there exists a collapsing perfect fluid that can be
described by the line element
\begin{equation}
\label{m1}
ds^2=dt^2-X^2dr^2-Y^2(d\theta^2+\sin^2\theta d\phi^2),
\end{equation}
where $X$ and $Y$ are functions of $r$ and $t$. Since we shall assume that
the source is dust, which always moves along geodesics, the system used in
(\ref{m1}) is considered comoving. For the exterior spacetime to $\Sigma$,
since
the fluid is not radiating, we have the Schwarzschild-de Sitter spacetime,
\begin{equation}
\label{m2}
ds^2=FdT^2-\frac{1}{F}dR^2-R^2(d\theta^2+\sin^2\phi d\phi^2),
\end{equation}
where $F$ is a function of $R$ given by
\begin{equation}
\label{ss}
F(R)=1-\frac{2M}{R}-\frac{\Lambda}{3}R^2,
\end{equation}
where $M$ is a constant and $\Lambda$ the cosmological constant.
In accordance with the Darmois junction conditions \cite{Darmois,Bonnor}
we
suppose that the first and second fundamental forms inherited by $\Sigma$
from the interior (\ref{m1}) and exterior (\ref{m2}) spacetimes are the
same. The
conditions are necessary and sufficient for a smooth matching.
The equations of $\Sigma$ may be written
\begin{eqnarray}
\label{c1}
r-r_{\Sigma}=0\; \mbox{in}\; V^-,\\
\label{c2}
R-R_{\Sigma}(T)=0\; \mbox{in}\; V^+,
\end{eqnarray}
where $V^-$ refers to the spacetime interior of $\Sigma$ and $V^+$ to the
spacetime exterior, and $r_{\Sigma}$ is a constant because $\Sigma$ is a
comoving surface forming the boundary of the dust.To apply the junction
conditions we must arrange that $\Sigma$ has the same parametrisation
whether it is considered as embedded in $V^+$ or in $V^-$.
We have for the metric (\ref{m1}) using (\ref{c1},\ref{c2}) on $\Sigma$,
\begin{equation}
\label{m3}
ds^2_{\Sigma}=dt^2-[Y(r_{\Sigma},t)]^2(d\theta^2+\sin^2\theta d\phi^2),
\end{equation}
We shall take $\xi^0=t$, $\xi^2=\theta$ and $\xi^3=\phi$ as the parameters
on $\Sigma$, hence the first fundamental form of $\Sigma$ can be written,
\begin{equation}
\label{fff}
g_{ij}d\xi^i d\xi^j,
\end{equation}
with latin indices ranging 0,2,3 and $g_{ij}$ the metric on $\Sigma$.
The metric (\ref{m2}) considering (\ref{m3}) becomes on $\Sigma$
\begin{equation}
ds^2_{\Sigma}=\left[F(R_{\Sigma})-\frac{1}{F(R_{\Sigma})}
\left(\frac{dR_{\Sigma}}{dT}\right)^2\right]dT^2-R^2_{\Sigma}(d\theta^2+
\sin^2\theta d\phi^2),
\end{equation}
where we assume
\begin{equation}
F(R_{\Sigma})-\frac{1}{F(R_{\Sigma})}\left(\frac{dR_{\Sigma}}{dT}\right)^2
>0,
\end{equation}
so that $T$ is a timelike coordinate. Now considering the continuity of
the
first fundamental form (\ref{fff}) from the metrics on $\Sigma$ (\ref{m3})
and (\ref{fff})
we get
\begin{eqnarray}
\label{rd}
R_{\Sigma}=Y(r_{\Sigma},t),\\
\label{td}
\left[F(R_{\Sigma})-\frac{1}{F(R_{\Sigma})}\left(\frac{dR_{\Sigma}}{dT}
\right)^2\right]^{1/2}dT=dt.
\end{eqnarray}
The second fundamental form of $\Sigma$ is
\begin{equation}
K_{ij}d\xi^i d\xi^j,
\end{equation}
where $K_{ij}$ is the extrinsic curvature given on the two sides by
\begin{equation}
K_{ij}^{\pm}=-n^{\pm}_{ij}\frac{\partial^2x^{\alpha}}{\partial\xi^i
\partial\xi^j}-n^{\pm}_{\alpha}\Gamma^{\alpha}_{\beta\gamma}\frac{\partial
x^{\beta}}{\partial\xi^i}\frac{\partial x^{\gamma}}{\partial\xi^j}.
\end{equation}
The Christoffel symbols $\Gamma^{\alpha}_{\beta\gamma}$ are to be
calculated
from the appropriate interior or exterior metrics, (\ref{m1}) or
(\ref{m2}),
$n^{\pm}_{\alpha}$ are the outward unit normals to $\Sigma$ in $V^-$ and
$V^+$ which come from (\ref{c1}) and (\ref{c2})
\begin{eqnarray}
n^-_{\alpha}=[0,X(r_{\Sigma},t),0,0],\\
n^+_{\alpha}=(-\dot{R}_{\Sigma},\dot{T},0,0),
\end{eqnarray}
where the dot stands for differentiation with respect to $t$, and
$x^{\alpha}$ refers to the equation of $\Sigma$ (\ref{c1}) and (\ref{c2}).
The nonzero
$K^{\pm}_{ij}$ are the following
\begin{eqnarray}
K^+_{22}=\csc^2\theta
K^+_{33}=\left(\frac{YY^{\prime}}{X}\right)_{\Sigma},\\
K^+_{00}=\left(\dot{R}\ddot{T}-\dot{T}\ddot{R}-\frac{F}{2}
\frac{dF}{dR}\dot{T}^3+\frac{3}{2F}
\frac{dF}{dR}\dot{T}\dot{R}^2\right)_{\Sigma},\\
K^+_{22}=\csc^2\theta K^+_{33}=(FR\dot{T})_{\Sigma}.
\end{eqnarray}
The continuity of the second fundamental form imposes
\begin{equation}
\label{jc}
K^+_{00}=0, \; K^-_{22}=K^+_{22},
\end{equation}
on the nonzero components of the extrinsic curvature. Now by considering
(\ref{jc}) with the conditions (\ref{rd},\ref{td}) and with (\ref{ss}) we
have
\begin{eqnarray}
\label{ej1}
(X\dot{Y}^{\prime}-\dot{X}Y^{\prime})_{\Sigma}=0,\\
\label{ej2}
M=\left(\frac{Y}{2}-\frac{\Lambda}{6}Y^3+\frac{Y}{2}\dot{Y}^2-\frac{Y}{2X^2}
Y^{\prime 2}\right)_{\Sigma}.
\end{eqnarray}
Let us now summarise the junction conditions of $\Sigma$. The necessary
and
sufficient conditions for smooth matching of metric (\ref{m1}), being
comoving, and
metric (\ref{m2}) are (\ref{rd},\ref{td}) and (\ref{ej1},\ref{ej2}).

\section{Field equations}

The spherical dust collapse with cosmological constant is described by
Einstein's field equations,
\begin{equation}
\label{fe}
R_{\alpha\beta}=8\pi\rho\left(u_{\alpha}u_{\beta}-
\frac{1}{2}g_{\alpha\beta}\right)-\Lambda g_{\alpha\beta},
\end{equation}
where $\rho$ is the energy density. Considering the line element
{\ref{m1}) and
numbering the coordinates $x^0=t$, $x^1=r$, $x^2=\theta$ and $x^3=\phi$ we
obtain for the components of (\ref{fe}), with the comoving four velocity
$u_{\alpha}=\delta^0_{\alpha}$,
\begin{eqnarray}
\label{eom1}
R_{00}=-\frac{\ddot{X}}{X}-2\frac{\ddot{Y}}{Y}=4\pi\rho-\Lambda,\\
\label{eom2}
R_{11}=\frac{\ddot{X}}{X}+2\frac{\dot{X}}{X}\frac{\dot{Y}}{Y}-\frac{2}{X^2}
\left(\frac{Y^{\prime\prime}}{Y}-\frac{X^{\prime}}{X}\frac{Y^{\prime}}{Y}
\right)=4\pi\rho+\Lambda,\\
\label{eom3}
R_{22}=\frac{\ddot{Y}}{Y}+\left(\frac{\dot{Y}}{Y}\right)^2+\frac{\dot{X}}{X}
\frac{\dot{Y}}{Y}-\frac{1}{X^2}\left[\frac{Y^{\prime\prime}}{Y}+
\left(\frac{Y^{\prime}}{Y}\right)^2-\frac{X^{\prime}}{X}
\frac{Y^{\prime}}{Y}-\left(\frac{X}{Y}\right)^2\right]\nonumber\\
=\csc^2\theta R_{33}=4\pi\rho+\Lambda,\\
\label{eom4}
R_{01}=-2\frac{\dot{Y}^{\prime}}{Y}+2\frac{\dot{X}}{X}\frac{Y^{\prime}}{Y}=0
.
\end{eqnarray}
Integrating (\ref{eom4}) we obtain
\begin{equation}
\label{ip}
X=\frac{Y^{\prime}}{W},
\end{equation}
where $W=W(r)$ is an arbitrary function of $r$. From (\ref{eom1},
\ref{eom2},\ref{eom3}) with (\ref{ip}) we
obtain
\begin{equation}
\label{28}
2\frac{\ddot{Y}}{Y}+\left(\frac{\dot{Y}}{Y}\right)^2+\frac{1-W^2}{Y^2}=
\Lambda,
\end{equation}
and after integration,
\begin{equation}
\label{bound}
\dot{Y}^2=W^2-1+2\frac{m}{Y}+\frac{\Lambda}{3}Y^2,
\end{equation}
where $m=m(r)$ is an arbitrary function of $r$. Substituting
(\ref{ip},\ref{bound}) into
either of (\ref{eom1},\ref{eom2},\ref{eom3}) we obtain
\begin{equation}
\label{dmf}
m^{\prime}=4\pi\rho Y^2Y^{\prime}.
\end{equation}
Integrating (\ref{dmf}) we obtain
\begin{equation}
\label{mf}
m(r)=4\pi\int^{r}_{0}\rho Y^2dY+m_0,
\end{equation}
where $m_0$ is an arbitrary constant. Since we want a finite distribution
of
matter at the origin $r=0$ we assume $m_0=0$.
From the field equation (\ref{eom4}) we see that the junction condition
(\ref{ej1}) is
identically satisfied. Substituting into the junction condition
(\ref{ej2})
equations (\ref{ss},\ref{ip},\ref{bound}) we get
\begin{equation}
M=m_{\Sigma}.
\end{equation}
From (\ref{ss}) we see that if $\Lambda=0$ the exterior spacetime becomes
the
Schwarzschild spacetime and $M$ is interpreted as the total energy inside
$\Sigma$ because of its Newtonian asymptotic behaviour.
To calculate the total energy ${\cal M}(r,t)$ up to a radius $r$ at a time
$t$ inside $\Sigma$ we use the definition of mass function
\cite{Misner,Cahill,Herrera} which is proportional to the component of the
Riemann tensor ${R^{23}}_{23}$, and is given for the metric (\ref{m1}) as
\begin{equation}
\label{function}
{\cal
M}(r,t)=\frac{1}{2}Y^3{R^{23}}_{23}=\frac{1}{2}Y\left[1-
\left(\frac{Y^{\prime}}{X}\right)^2+\dot{Y}^2\right].
\end{equation}
Using (\ref{ip},\ref{bound}) into (\ref{function}), we get
\begin{equation}
\label{mf2}
{\cal M}(r,t)=m(r)+\frac{\Lambda}{6}Y^3(r,t).
\end{equation}
The quantity $m(r)$ can be interpreted as energy due to the energy density
$\rho(r,t)$, given by (\ref{mf}), and since it is measured in a comoving
frame $m$
is only $r$ dependent. If we give a perfect fluid intepretation to the
cosmological constant $\Lambda$ like
\begin{equation}
\Lambda=8\pi\mu,
\end{equation}
where $\mu$ is a constant energy density and with a pressure $p=-\mu$,
then
the second term in the right hand side of (\ref{mf2}) becomes,
\begin{equation}
\frac{\Lambda}{6}Y^3=\frac{4\pi}{3}\mu Y^3.
\end{equation}
Since the related cosmological fluid is not comoving with the frame inside
$\Sigma$, its associated energy is time dependent. Its contribution to the
total energy is positive or negative accordingly if $\Lambda>0$ or
$\Lambda<0$.
Another expression for the total energy inside $\Sigma$, for slow
collapse,
has been proposed by Tolman \cite{Tolman1} and Whittaker \cite{Whittaker}
(for a discussion about its physical meaning see \cite{Herrera}). That
expression reduces, in our case,to (\ref{mf}).
When shells of matter cross each other, there are shell crossing
singularities \cite{Eardley} and to avoid them we need that the proper
radius increases with the coordinate $r$, hence we require from (\ref{ip})
\begin{equation}
X(r,t)>0.
\end{equation}
Since we assume $\rho(r,t)>0$ then (\ref{dmf}) gives
\begin{equation}
m^{\prime}\geq0,
\end{equation}
which means that the energy density increases with $r$.

\section{Dust solution with $W(r)=1$}

From now on we consider only the case $\Lambda>0$ and the
assumption
\begin{equation}
\label{mb}
W(r)=1.
\end{equation}
From (\ref{ip}), with (\ref{mb}), both the radii of the collapsing sphere,
calculated from
$g_{11}$ and from $g_{33}$ (which follows from the perimeter divided by
$2\pi$), are the same. The condition (\ref{mb}) in the absence of
cosmological constant
is the marginally bound condition, limiting the situations where the shell
is bounded from
those it is unbounded. In the presence of a cosmological constant, the
situation is more complex,
and $W(r)=1$ leads to an unbounded shell. When $W(r) < 1$, there is
unbounded shell for
$\Lambda > \Lambda_c$ and bounded for $\Lambda < \Lambda_c$, $\Lambda_c$
being
the root of (\ref{bound}).
Then with (\ref{mb}) we obtain from (\ref{ip},\ref{bound}),
\begin{eqnarray}
\label{s1}
Y(r,t)=\left(\frac{6m}{\Lambda}\right)^{1/3}\sinh^{2/3}\alpha(r,t),\\
\label{s2}
X(r,t)=\left(\frac{6m}{\Lambda}\right)^{1/3}\left[\frac{m^{\prime}}{3m}
\sinh\alpha(r,t)\right. \nonumber\\
\left. +\left(\frac{\Lambda}{3}\right)^{1/2}t^{\prime}_0
\cosh\alpha(r,t)
\right]\sinh^{-1/3}\alpha(r,t),
\end{eqnarray}
where
\begin{equation}
\alpha(r,t)=\frac{\sqrt{3\Lambda}}{2}[t_0(r)-t],
\end{equation}
and $t_0(r)$ is an arbitrary function of $r$. For $t=t_0(r)$ we have
$Y(r,t)=0$ which is the time when the matter shell $r=$constant hits the
physical singularity. Taking the limit $\Lambda\rightarrow 0$ of
(\ref{s1},\ref{s2}) we
reobtain the Tolman-Bondi solution \cite{Tolman,Bondi}
\begin{eqnarray}
\lim_{\Lambda\rightarrow
0}Y(r,t)=\left[\frac{9m}{2}(t_0-t)^2\right]^{1/3},\\
\lim_{\Lambda\rightarrow 0}X(r,t)=\frac{m^{\prime}(t_0-t)+2mt^{\prime}_0}
{[6m^2(t_0-t)]^{1/3}}.
\end{eqnarray}

\section{Apparent horizons}

The apparent horizon is formed when the boundary of trapped two spheres
are
formed. We search for two spheres whose outward normals are null, which
give
for (\ref{m1}),
\begin{equation}
\label{45}
g^{\mu\nu}Y_{,\mu}Y_{,\nu}=-{\dot
Y}^2+\left(\frac{Y^{\prime}}{X}\right)^2=0.
\end{equation}
Considering (\ref{ip},\ref{bound}) we have from (\ref{45}),
\begin{equation}
\label{ea}
\Lambda Y^3-3Y+6m=0
\end{equation}
where the solutions for $Y$ give the apparent horizons. For $\Lambda=0$ we
have the Schwarzschild horizon $Y=2m$, and for $m=0$ we have the de Sitter
horizon $Y=\sqrt{3/\Lambda}$. For $3m<1/\sqrt\Lambda$ there are two
horizons,
\begin{eqnarray}
\label{root1}
Y_1=\frac{2}{\sqrt\Lambda}\cos\frac{\varphi}{3},\\
\label{root2}
Y_2=-\frac{1}{\sqrt\Lambda}\left(\cos\frac{\varphi}{3}-\sqrt 3
\sin\frac{\varphi}{3}\right),
\end{eqnarray}
where $\varphi$ is given by
\begin{equation}
\label{d}
\cos\varphi=-3m\sqrt\Lambda.
\end{equation}
If $m=0$ we have $Y_2=0$ and $Y_1=\sqrt{3/\Lambda}$, then we call $Y_1$
the
cosmological horizon generalized when $m\neq 0$, and $Y_2$ the black hole
horizon generalized when $\Lambda\neq 0$ \cite{Hayward}.
For $3m=1/\sqrt\Lambda$ (\ref{root1}) and (\ref{root2}) coincide and there
is only one
horizon,
\begin{equation}
\label{mass}
Y=\frac{1}{\sqrt\Lambda}
\end{equation}
In general, the range for $Y_1$ and $Y_2$ is given by
\begin{equation}
\label{leq}
0 \leq Y_2 \leq \frac{1}{\sqrt{\Lambda}} \leq Y_1 \leq
\sqrt{\frac{3}{\Lambda}}.
\end{equation}
The biggest amount of mass $m$ for the formation of an
apparent black hole horizon is given by (\ref{mass}), attaining at that
stage its
largest proper area $4\pi/Y^2(r,t)=4\pi/\Lambda$. The cosmological horizon
has an area
between $4\pi/\Lambda$ up to $12\pi/\Lambda$.
For $3m>1/\sqrt\Lambda$ there are no horizons.
The time for the formation of the apparent horizon, for $W(r)=1$, can be
obtained from (\ref{s1})
with (\ref{ea}) giving
\begin{equation}
t_n=t_0-\frac{2}{\sqrt{3\Lambda}}\arg\sinh\left(\frac{Y_n}{2m}-1
\right)^{1/2},
\end{equation}
where $Y_n$ stands for either of the values (\ref{root1},\ref{root2}) or
(\ref{mass}). In the limit
when $\Lambda=0$ it is reobtained the Tolman-Bondi result \cite{Eardley},
\begin{equation}
\label{ta}
t_{AH}=t_0-\frac{4}{3}m.
\end{equation}
\par
From (\ref{leq}) we have that both apparent horizons, black hole and
cosmological,
precede the singularity $t=t_0$ by an amount of comoving time
$(2/\sqrt{3\Lambda})\arg\sinh\sqrt{(Y_n/2m)-1}$. From (\ref{leq}) we can
write
\begin{equation}
\label{Y}
\frac{Y_n}{2m}=\cosh^2\alpha_n.
\end{equation}
Since, from (\ref{leq}), always $Y_1\geq Y_2$, we have also from
(\ref{ta})
$\alpha_1\geq\alpha_2$ or
$t_1\leq t_2$, which means that the cosmological horizon always precedes
the
black hole horizon. To see how $t_n$ behaves as a function of $m$, we
first
calculate, using (\ref{root1},\ref{root2},\ref{d}),
\begin{eqnarray}
\label{Y1}
\frac{d(Y_1/2m)}{dm}=\frac{1}{m}\left(-\frac{\sin\varphi/3}{\sin\varphi}+
\frac{3\cos\varphi/3}{\cos\varphi}\right)<0,\\
\label{Y2}
\frac{d(Y_2/2m)}{dm}=\frac{1}{m}\left[-\frac{\sin(\varphi+4\pi)/3}
{\sin\varphi}+\frac{3\cos(\varphi+4\pi)/3}{\cos\varphi}\right]>0.
\end{eqnarray}
Defining
\begin{equation}
\tau_n=t_0-t_n,
\end{equation}
we have from (\ref{Y}),
\begin{equation}
\label{tau}
\frac{d\tau_n}{d(Y_n/2m)} =
\frac{1}{\sqrt{\Lambda}\sinh\alpha_n\cosh\alpha_n}.
\end{equation}
Now considering (\ref{Y1}) and (\ref{root1},\ref{tau}) we have
\begin{eqnarray}
\label{tau1}
\frac{d\tau_1}{dm}=\frac{d\tau_1}{d(Y_1/2m)}\frac{d(Y_1/2m)}{dm}\nonumber\\
=\frac{1}{m\sqrt{3\Lambda}\sinh\alpha_1\cosh\alpha_1}\left(-
\frac{\sin\varphi/3}{\sin\varphi}+
\frac{3\cos\varphi/3}{\cos\varphi}\right)<0.
\end{eqnarray}
\par
From (\ref{tau1}) we have that $\tau_1$ decreases when $m$ increases,
which
means
that the time interval between the formation of the singularity and the
cosmological horizon diminishes while $m$ increases.
While considering (\ref{root2}) and (\ref{Y2}) we have
\begin{eqnarray}
\label{tau2}
\frac{d\tau_2}{dm}=\frac{1}{m\sqrt{3\Lambda}\sinh\alpha_2\cosh\alpha_2}\nonumber\\
\times\left[-\frac{\sin(\varphi+4\pi)/3}{\sin\varphi}+
\frac{3\cos(\varphi+4\pi)/3}
{\cos\varphi}\right]>0.
\end{eqnarray}
Hence from (\ref{tau2}) we have that $\tau_2$ decreases when $m$
increases,
meaning
that the time interval between the formation of the singularity and the
black hole horizon increases while $m$ increases.
\par
The nature of the surface of the two physical horizons, the cosmological
and
the
black hole horizons, can be obtained calculating the induced metric on
them.
This can be done using
\begin{equation}
dt_n = \biggr(t_0' - \frac{2}{\sqrt{3\Lambda}}\alpha'_n\biggl)dr
\end{equation}
and introducing the roots for $Y_n$. For the cosmological horizon, we
obtain
that
it is always past timelike. On the other hand, the nature of the black
hole
horizon
depends crucially on the value of $t_0'$. Considering $m' > 0$, we obtain,
\begin{eqnarray}
0&\leq&t_0'\leq\frac{m'}{6m}\frac{Y_2(1+\Lambda Y^2_2)}{1-\Lambda
Y^2_2}\quad
\mbox{(past timelike)},\\
0&<&t_0'=\frac{m'}{6m}\frac{Y_2(1+\Lambda Y^2_2)}{1-\Lambda Y^2_2}\quad
\mbox{(past null)},\\
0&<&\frac{m'}{6m}\frac{Y_2(1+\Lambda Y^2_2)}{1-\Lambda Y^2_2}<t_0' \quad
\mbox{(spacelike)}.
\end{eqnarray}
In the limit $\Lambda\rightarrow0$, we find the results of \cite{Eardley}.

\section{Conclusions}

In the collapse of a dust matter in a spherically symmetric spacetime,
a black hole is inevitably formed. The
singularity is preceded by the formation of an apparent horizon by a lapse
of time $\Delta t = (4/3)m$, where $m$ is the black hole mass. The
dimension,
consequently the mass of the black hole, is not limited in principle.
In this analysis, it is employed the Tolman-Bondi metric which is
asymptotically
flat. In this case, the mass of the configuration is the same either if we
use
the Cahill-McVittie or the Tolman-Whittaker criteria.
\par
The introduction of a cosmological constant changes this scenario in many
ways.
There are now three apparent horizons instead of one. One of these
apparent
horizon
is not physical, while the other two represent the black hole horizon and
the
cosmological horizon. The spacetime is no more asympotically flat. The
cosmological
constant affects the lapse of time between the formation of the apparent
horizon and
the singularity.
\par
The two main consequences of introducing the cosmological constant,
however,
concerns the mass definition and the dimension of the black hole.
Using the different definitions of mass in a gravitating system in a
curved
background can lead to a mass that includes or not the
energy associated with the cosmological constant. Moreover, a black hole
with
a mass that exceeds $m = 1/\sqrt{\Lambda}$ can not be formed. This result
has already been obtained in a static spherically symmetric configuration;
but it remains when we introduce dynamics in the system.
\par
In general, the introduction of a cosmological constant slows down the
collapse
process. The cosmological constant leads, when we think in terms of a
Newtonian
potential, to a repulsive term. Taking the $g_{00}$ in the case of a
static
symmetric
spacetime in presence of a cosmological constant, and considering that
$g_{00} = 1 - 2\phi(R)$, where $\phi = (m/R) + (1/6)\Lambda R^2$, so that
the Newtonian force is given by
\begin{equation}
F(R)= - \frac{m}{R^2} + \frac{\Lambda}{3}R \quad .
\end{equation}
For $R = 1/\sqrt{\Lambda}$ and $m = (1/3)\sqrt{\Lambda}$ (65) leads to $F
=
0$. Mass and radius, larger
than those ones, the force becomes repulsive, while for smaller values of
mass and radius, the force becomes attractive, leading to the collapse and
formation
of a singularigy. In this sense, the limitation of the size of the black
hole
can be understood in terms of this competition between repulsive and
attractive
forces, where the first comes from the cosmological constant term.
\par
The slow down of the collapse process can be verified from the expression
\begin{equation}
\ddot Y = - \frac{m}{Y^2} + \frac{\Lambda}{3}Y,
\end{equation}
obtained from (\ref{28},\ref{bound}).
This expression reproduces the Newtonian model formulated
before:
the presence of a cosmological constant diminishes the acceleration in the
collapse process,
and for $Y\geq(3m/{\Lambda})^{1/3}$, there is no collapse at all, since
the
acceleration becomes
positive.
\newline
\vspace{0.5cm}
\newline
{\bf Acknowledgements:} We thank CAPES and CNPq (Brazil) for financial
support.


\begin{thebibliography}{99}
\bibitem{Weinberg}Weinberg S 1996 {\it Theories of the cosmological
constant} astro-ph/9610044
\bibitem{Cohn}Cohn J D {\it Living with Lambda} astro-ph/9807128
\bibitem{Riess}Riess A G et al. {\it Observational evidence from
supernovae
for an
accelerating Universe and a cosmological constant} astro-ph/9805201
\bibitem{Perlmutter}Perlmutter et al. 1998 {\it Nature} {\bf 391} 51
\bibitem{Garfinkle}Garfinkle D and Vuille C 1991 {\it Gen. Rel. Grav.}
{\bf
23} 471
\bibitem{Hayward}Hayward S A, Shiromizu T and Nakao K 1994 {\it Phys. Rev.
D} {\bf 49} 5080
\bibitem{Darmois}Darmois G 1927 {\it M\'{e}morial des Sciences
Math\'{e}matiques} (Paris: Gauthier-Villars) Fasc. 25
\bibitem{Bonnor}Bonnor W B and Vickers P A 1981 {\it Gen. Rel. Grav.} {\bf
13} 29
\bibitem{Misner}Misner C and Sharp D 1964 {\it Phys. Rev.} {\bf 136} B571
\bibitem{Cahill}Cahill M and McVittie G 1970 {\it J. Math. Phys.} {\bf 11}
1382
\bibitem{Herrera}Herrera L and Santos N O 1995 {\it Gen. Rel. Grav.} {\bf
27} 1071
\bibitem{Eardley}Eardley D M and Smarr L 1979 {\it Phys. Rev. D} {\bf 19}
2239
\bibitem{Tolman}Tolman R C 1934 {\it Proc. Nat. Acad. Sci. USA} {\bf 20}
164
\bibitem{Bondi}Bondi H 1947 {\it Mon. Not. R. Astron. Soc.} {\bf 107} 410
\bibitem{Tolman1}Tolman R C 1930 {\it Phys. Rev.} {\bf 35} 875
\bibitem{Whittaker}Whittaker E T 1935 {\it Proc. Roy. Soc. Lond.} {\bf
A149}
384
\end{thebibliography}
\end{document}